\documentclass[11pt]{article}

\usepackage{amsmath}
\usepackage{graphicx}
\usepackage{amsfonts}
\usepackage{amssymb}
\usepackage{epsfig}
\usepackage{color}
\usepackage{psfrag}
\usepackage{epstopdf}

\newcommand{\EQ}{\begin{equation}}
\newcommand{\EN}{\end{equation}}
\newcommand{\be}{\begin{equation}}
\newcommand{\ee}{\end{equation}}
\newcommand{\bea}{\begin{eqnarray}}
\newcommand{\eea}{\end{eqnarray}}

\begin{document} \setcounter{page}{0}
\topmargin 0pt
\oddsidemargin 5mm
\renewcommand{\thefootnote}{\arabic{footnote}}
\newpage
\setcounter{page}{0}
\topmargin 0pt
\oddsidemargin 5mm
\renewcommand{\thefootnote}{\arabic{footnote}}
\newpage
\begin{titlepage}
\begin{flushright}
SISSA 57/2011/EP \\
\end{flushright}
\vspace{0.5cm}
\begin{center}
{\large {\bf Crossing probability and number of crossing clusters in off-critical percolation}}\\
\vspace{1.8cm}
{\large Gesualdo Delfino and Jacopo Viti}\\
\vspace{0.5cm}
{\em SISSA -- Via Bonomea 265, 34136 Trieste, Italy}\\
{\em INFN sezione di Trieste}\\
\end{center}
\vspace{1.2cm}

\renewcommand{\thefootnote}{\arabic{footnote}}
\setcounter{footnote}{0}

\begin{abstract}
\noindent
We consider two-dimensional percolation in the scaling limit close to criticality and use integrable field theory to obtain universal predictions for the probability that at least one cluster crosses between opposite sides of a rectangle of sides much larger than the correlation length and for the mean number of such crossing clusters.
\end{abstract}
\end{titlepage}

\newpage
\section{Introduction}
Remarkable results have been obtained in the last two decades for crossing clusters in two-dimensional percolation at its critical point $p_c$. In particular, following the numerical study of \cite{LPPS-A}, Cardy \cite{Cardy92} used conformal field theory to derive an exact formula for the crossing probability $P_v$ that, in the continuum limit, at least one cluster spans between the horizontal sides of a rectangle\footnote{See \cite{Watts} for the probability of simultaneous horizontal and vertical crossing. At $p_c$ results for the rectangular geometry are connected by conformal symmetry to other simply connected domains \cite{Cardy92,LPS-A}.}, a result later proved rigorously by other methods \cite{S.Smirnov}. Cardy also determined the mean number $\bar{N}_v$ of crossing clusters \cite{Cardy00, CardyLP}.

For a rectangle of width $L$ and height $R$, as a consequence of scale invariance both $P_v$ and $\bar{N}_v$ depend at $p_c$ only on the aspect ratio $R/L$. In this paper we consider these quantities in the scaling limit close to $p_c$, where, due to the presence of a finite correlation length $\xi$ (much larger than the lattice spacing), they separately depend on $L/\xi$ and $R/\xi$. We consider the limit $L\gg\xi$ and use boundary integrable field theory to determine the mean number of vertically crossing clusters, i.e. the clusters which span between the sides of the rectangle separated by the distance $R$, in the limit $R\gg\xi$. The result we obtain below $p_c$ is given in (\ref{spanning_clusters_result}), (\ref{phi}). On the other hand, we can observe that for $R\to\infty$ below $p_c$ vertical crossing becomes extremely rare, so that $P_v\equiv\text{Prob}(N_v>0)\sim\text{Prob}(N_v=1)\sim\bar{N}_v$; from the leading term in (\ref{phi}) we then obtain for the vertical crossing probability in the scaling limit below $p_c$ the universal result
\EQ
\label{P}
P_v(L,R)\sim A~\frac{L}{\xi}\,\text{e}^{-R/\xi}\,,\hspace{1cm}L\gg\xi\,,\hspace{.3cm}R\gtrsim L\,,
\EN
where
\EQ
A=\frac{1}{2}\,(3-\sqrt{3})\,.
\label{A}
\EN
The correlation length $\xi$ we refer to is defined by the decay of the probability $P_2(r)$ that two points separated by a distance $r$ are in the same finite cluster:
\EQ
P_2(r)\propto r^{-a}\,e^{-r/\xi}\,,\hspace{1cm}r\to\infty\,,
\label{xi1}
\EN
with $a=1/2$ below $p_c$ and $a=2$ above $p_c$ \cite{DVC}; $\xi$ is related to the mass $m$ appearing in (\ref{phi}) and throughout the paper as 
\EQ
\xi=\left\{\begin{array}{ccc}1/m\,, & \hspace{.3cm}p<p_c\,,
  \\  1/2m\,, & \hspace{.3cm}p>p_c\,.\end{array}\right.
\label{xi}
\EN
We will also give a direct derivation of (\ref{P}) which also yields the next term, given by (\ref{U}), in the large $R$ expansion. The corresponding results in the scaling limit above $p_c$ are given in (\ref{Nv+}) and (\ref{Ptilde}). 

It has been previously known \cite{BW,HA} that for $R=L\gg\xi$ the crossing probability is a function of $L|p-p_c|^\nu$ which decays exponentially to zero below $p_c$ and to one above, a feature investigated numerically in \cite{HA,NZ,ONS,Vasilyev}. 

We will start the analysis recalling in the next section the relation with the $q$-state Potts model and explaining how the latter is described in the scaling limit within the framework of boundary integrable field theory. In section~3 we take the limit $q\to 1$ relevant for percolation and determine the quantities of our interest.

\begin{figure}[t]
\begin{center}
\includegraphics[height=5cm]{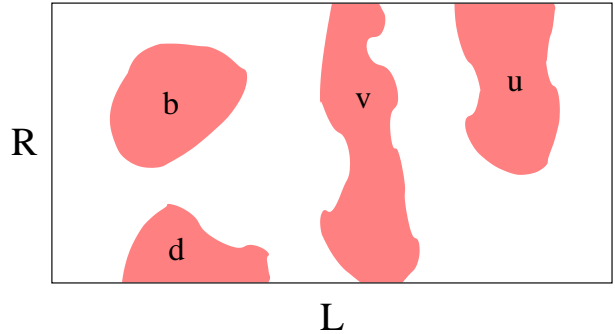}
\caption{The different types of clusters which, depending on boundary conditions, determine the $q$-dependence of the Potts partition functions (\ref{Zaa}-\ref{Zfa}).}
\label{cross_cluster}
\end{center}
\end{figure}

\section{Mapping to the Potts model and field theory}
It is well known that an efficient theoretical approach to random percolation is to see it as a limiting case of the $q$-state Potts model defined by the lattice Hamiltonian \cite{Potts,Wu}
\EQ
{\cal H}=-J\sum_{\langle i,j\rangle}\delta_{s_i,s_j}\,,\hspace{1cm}s_i=1,\ldots,q\,.
\label{H}
\EN
The partition function $\sum_{\{s_i\}}e^{-{\cal H}}$ admits the expansion \cite{FK} $\sum_{\cal G}p^{n}(1-p)^{\bar{n}}q^{N_{clusters}}$ over bond configurations ${\cal G}$, where $p=1-e^{-J}$,  $n$ is the number of bonds in ${\cal G}$, $\bar{n}$ the complement to the total number of edges, and each cluster formed by adjacent bonds contributes a factor $q$ corresponding to the number of colors\footnote{Different values of the Potts spins can conveniently be associated to different colors. The Hamiltonian (\ref{H}) is invariant under permutations of the colors.} it can take; for $q\to 1$ configurations are weighted as required for bond percolation. The above expression for the partition function holds as it is for free (f) boundary conditions. If instead the color of the spins on a boundary is fixed to be $\alpha$, the clusters touching that boundary can take only the color $\alpha$ and do not contribute any factor $q$ to the weight. In particular, if we denote by $Z_{ud}^{lr}$ the Potts partition function on a rectangle with boundary conditions $u$, $d$, $l$ and $r$ on the upper, lower, left and right boundary, respectively, as already noted in \cite{Cardy00, CardyLP} we have
\begin{align}
\label{Zaa}
&Z_{\alpha\alpha}^{ff}=\sum_{\mathcal{G}}p^{n}(1-p)^{\bar{n}}q^{N_b},\qquad\qquad Z_{ff}^{ff}=\sum_{\mathcal{G}}p^{n}(1-p)^{\bar{n}}q^{N_b+N_v+N_u+N_d},\\
&Z_{\alpha f}^{ff}=\sum_{\mathcal{G}}p^{n}(1-p)^{\bar{n}}q^{N_b+N_d},\qquad ~Z_{f\alpha}^{ff}=\sum_{\mathcal{G}}p^{n}(1-p)^{\bar{n}}q^{N_b+N_u},
\label{Zfa}
\end{align}
where $N_b$ is the number of clusters which do not touch the horizontal boundaries, $N_{u}$ ($N_d$) the number of clusters which touch the upper (lower) but not the lower (upper) boundary, and $N_v$ the number of those touching both horizontal boundaries (see Fig.~\ref{cross_cluster}). It follows that the mean number of vertically crossing clusters can be written as
\EQ
\bar{N}_v=\lim_{q\rightarrow 1}\partial_q\log\frac{Z_{ff}^{ff}Z_{\alpha\alpha}^{ff}}{Z_{\alpha f}^{ff}Z_{f\alpha}^{ff}}\,.
\label{Nc}
\EN
Since boundary conditions on Potts spins loose physical meaning as $q\to 1$ and sites do not interact in random percolation, (\ref{Nc}) gives the mean number of clusters spanning between the horizontal sides of a rectangular window within the infinite plane on which the percolative transition actually takes place.

Let us begin our field theoretical considerations for the scaling limit considering an infinitely long horizontal strip of height $R$.  With imaginary time running upwards, the partition functions on the strip can be written as
\EQ
Z_{ud}^{lr}=\langle B_u|\text{e}^{-RH}|B_d\rangle_{l,r}\,,
\label{Z_ud}
\EN
where $H$ is the Hamiltonian of the quantum system living in the infinite horizontal dimension, $|B_{d,u}\rangle$ are boundary states specifying initial and final conditions, and the vertical boundary conditions at infinity have the role of selecting the states which can propagate between the horizontal boundaries. Integrability of the scaling Potts model \cite{CZ} allows us to work in the framework of integrable field theories for which the bulk dynamics is entirely specified by the Faddeev-Zamolodchikov commutation rules (see e.g. \cite{Smirnov})
\begin{align}
\label{FZ1}
&A^{\dagger}_{i}(\theta_1)A^{\dagger}_{j}(\theta_2)=S_{ij}^{i'j'}(\theta_1-\theta_2)A^{\dagger}_{j'}(\theta_2)A^{\dagger}_{i'}(\theta_1)\,,\\
\label{FZ2}
&A^{i}(\theta_1)A_j^{\dagger}(\theta_2)=S^{j'i}_{ji'}(\theta_2-\theta_1)A_{j'}^{\dagger}(\theta_2)A^{i'}(\theta_1)+2\pi\delta^{i}_{j}\delta(\theta_1-\theta_2)\,,
\end{align}
where  $A_{i}^{\dagger}(\theta)$ and  $A^{i}(\theta)$ are creation and annihilation operators for a particle of species $i$ with  rapidity\footnote{Energy and momentum of a particle with mass $m$ are given by $(e,p)=(m\cosh\theta,m\sinh\theta)$.} $\theta$, and $S^{i'j'}_{ij}(\theta)$ are two-body scattering amplitudes satisfying, in particular, unitarity
\EQ
S^{i'j'}_{ij}(\theta)S^{i''j''}_{i'j'}(-\theta)=\delta^{i''}_{i}\delta^{j''}_{j}
\EN
and crossing symmetry
\EQ
S^{i'j'}_{ij}(\theta)=S^{j'\bar{i}}_{j\bar{i}'}(i\pi-\theta).
\EN
Generic boundary states $|B_a\rangle$ can be written as superpositions of asymptotic states of the particles created by $A_{i}^{\dagger}$, with vanishing total momentum in order to preserve horizontal translation invariance. The additional constraints coming from the requirement that a boundary condition preserves  integrability were discovered in \cite{GZ94}. In particular, particles carrying momentum can only appear in pairs with vanishing total momentum. On the other hand, for the case of our interest of a theory satisfying
\EQ
\label{fermionic_stat}
S_{ij}^{i'j'}(0)=(-1)\,\delta_i^{j'}\delta_{j}^{i'}\,,
\EN
states containing $k\geq 2$ particles of zero momentum are forbidden by (\ref{FZ1}). Finally $|B_a\rangle$ takes the form
\begin{equation}
\label{b_state}
|B_a\rangle=\mathcal{S}_a\exp\Bigl[\frac{1}{2}\int_{-\infty}^{\infty}\frac{d\theta}{2\pi}
~\mathcal{P}_a(\theta)\Bigr]|\Omega\rangle\,,
\end{equation}
where $|\Omega\rangle$ is the vacuum state and
\begin{align}
&\mathcal{S}_a=1+\tilde{g}_a^i A_i^{\dagger}(0)\,,\\
&\mathcal{P}_a(\theta)=K_a^{ij}(\theta)A_i^{\dagger}(-\theta)A_j^{\dagger}(\theta)\,.
\end{align}
The boundary pair emission amplitudes $K_a^{ij}(\theta)$ satisfy equations involving the bulk amplitudes $S_{ij}^{i'j'}(\theta)$, and the constants $\tilde{g}_a^i$ follow from the relations\footnote{See \cite{BPT07,DPTW00} for the factor 1/2 in (\ref{gtilde}).}
\EQ
2i\,\mbox{Res}_{\theta=0}K_a^{ij}(\theta)=g_a^ig_a^j\,,
\EN
\EQ
\tilde{g}_a^i=\frac{g_a^i}{2}\,.
\label{gtilde}
\EN
The exponential form of the boundary state  is a consequence of the boundary Yang-Baxter equations, which give in particular $[\mathcal{P}_a(\theta), \mathcal{P}_a(\theta')]=[\mathcal{P}_a(\theta),\mathcal{S}_a]=0$ \cite{GZ94}.

\begin{figure}[t]
\begin{center}
\includegraphics[height=4cm]{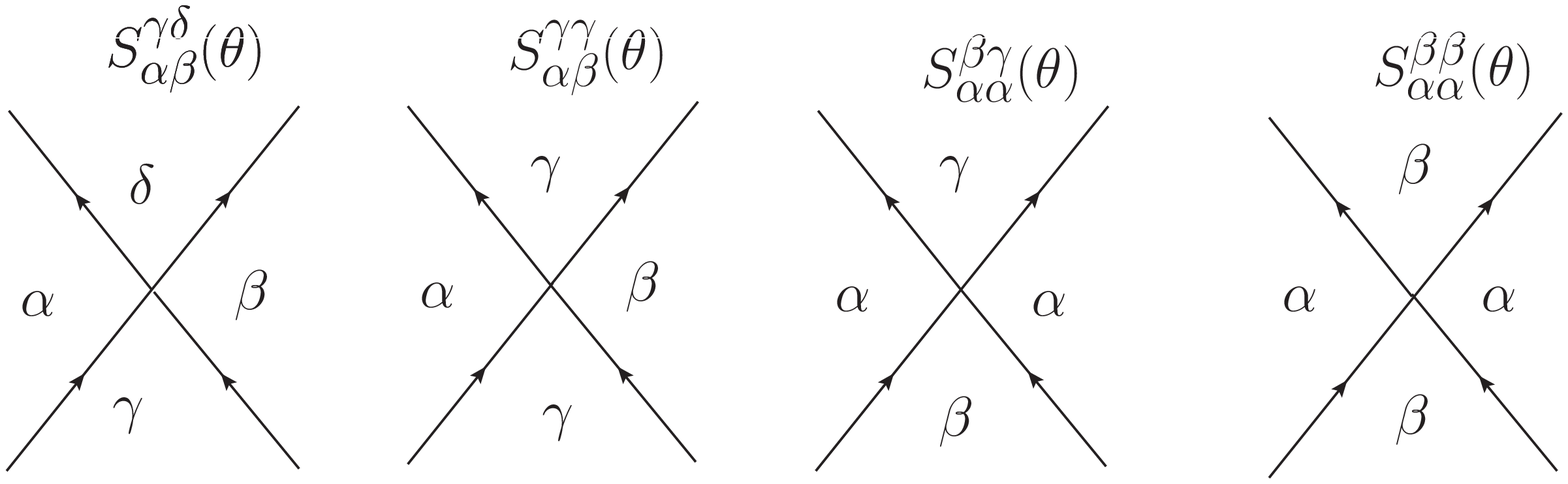}
\caption{The four inequivalent kink-kink scattering amplitudes in Potts field theory (different indices denote different colors).}
\label{fig_amplitudes}
\end{center}
\end{figure}

We are now ready to use this formalism to evaluate the Potts partition functions entering (\ref{Nc}). The Potts field theory, i.e. the integrable field theory which describes the scaling limit of the Potts model in two dimensions, was solved exactly in \cite{CZ} in the language of the spontaneously broken phase above $J_c$, in which the elementary excitations are kinks $A^{\dagger}_{\beta\alpha}(\theta)|\Omega_{\alpha}\rangle$ interpolating between degenerate ferromagnetic vacua $|\Omega_{\alpha}\rangle$ and $|\Omega_{\beta}\rangle$ with different color. The vacua satisfy $\langle\Omega_\alpha|\Omega_\beta\rangle=\delta_{\alpha\beta}$ and the admissible multi-kink states have the form
\EQ
A^{\dagger}_{\alpha_{n+1}\alpha_n}(\theta_n)\ldots A^{\dagger}_{\alpha_3\alpha_2}(\theta_2)A^{\dagger}_{\alpha_2\alpha_1}(\theta_1)|\Omega_{\alpha_1}\rangle\,.
\label{states}
\EN
For these topological excitations the Faddeev-Zamolodchikov commutation relations take the form,
\begin{align}
\label{FZ1kinks}
&A_{\alpha\beta}^{\dagger}(\theta_1)A_{\beta\gamma}^{\dagger}(\theta_2)=\sum_{\delta}S_{\alpha\gamma}^{\beta\delta}(\theta_1-\theta_2)A_{\alpha\delta}^{\dagger}(\theta_2)A_{\delta\gamma}^{\dagger}(\theta_1)\,,\\
\label{FZ2kinks}
&A_{\alpha\beta}(\theta_1)A_{\beta\gamma}^{\dagger}(\theta_2)=\sum_{\delta}S_{\beta\delta}^{\gamma\alpha}(\theta_2-\theta_1)A_{\alpha\delta}^{\dagger}(\theta_2)A_{\delta\gamma}(\theta_1)+2\pi\delta_{\alpha\gamma}\delta(\theta_1-\theta_2)\,,
\end{align}
where invariance of the theory under permutations of the colors allows for the four inequivalent  scattering amplitudes represented in Fig. \ref{fig_amplitudes}; they are given explicitly in \cite{CZ} and obey (\ref{fermionic_stat}) in the form
\EQ
\label{fermionic_kink}
S_{\alpha\gamma}^{\beta\delta}(0)=(-1)\,\delta_{\beta\delta}\,.
\EN

Both fixed (to a color $\alpha$) and free (f) boundary conditions are integrable and the corresponding pair emission amplitudes were determined in \cite{Chim}. They determine the boundary states $|B_\alpha\rangle$ and $|B_f\rangle$ in the form that we now specify; in the following $q$ will be parameterized as $q=4\sin^2\bigl(\frac{\pi\lambda}{3}\bigr)$, so that $\lambda\to 1/2$ corresponds to the  percolation limit. For fixed boundary conditions we have
\begin{equation}
|B_\alpha\rangle=\exp\Bigl[\frac{1}{2}\int_{-\infty}^{\infty}\frac{d\theta}{2\pi}
~\mathcal{P}_\alpha(\theta)\Bigr]|\Omega_\alpha\rangle\,,
\label{Pa}
\end{equation}
with
\EQ
\mathcal{P}_{\alpha}(\theta)=K_0(\theta)\sum_{\beta\not=\alpha}A^{\dagger}_{\alpha\beta}(-\theta)A^{\dagger}_{\beta\alpha}(\theta)\,,
\EN
\begin{align}
\label{K_fixed}
&K_0(\theta)=i\tanh\Bigl(\frac{\theta}{2}\Bigr)\exp\Biggl[-\int_{0}^{\infty}\frac{dt}{t}\frac{n_{\lambda}(t)}{2\cosh t}\sinh\Bigl(t-\frac{2\theta t}{i\pi}\Bigr)\Biggr],\\
&\quad n_{\lambda}(t)=\frac{\sinh\bigl(\frac{t}{6}+\frac{t}{2\lambda}\bigr)-\sinh\bigl(\frac{3t}{2}-\frac{t}{2\lambda}\bigr)}{\sinh\bigl(\frac{t}{2\lambda}\bigr)\cosh\bigl(\frac{t}{2}\bigr)}\,;
\end{align}
the integral in (\ref{K_fixed}) is convergent for $1/2\leq\lambda<1$.
$K_0(\theta)$ satisfies the boundary ``cross-unitarity'' relation
\EQ
\label{crossing_fixed}
K_0(\theta)=\bigl[S_{\alpha\alpha}^{\beta\beta}(2\theta)+(q-2)S_{\alpha\alpha}^{\beta\gamma}(2\theta)\bigr]K_0(-\theta)\,,
\EN
which together with (\ref{fermionic_kink}) implies $K_0(0)=0$, as already apparent from (\ref{K_fixed}); the absence of a pole at $\theta=0$ explains $\mathcal{S}_{\alpha}=1$.

For free boundary conditions we have instead
\begin{equation}
|B_f\rangle=\sum_\alpha{\cal S}_f^\alpha\,\exp\Bigl[\frac{1}{2}\int_{-\infty}^{\infty}\frac{d\theta}{2\pi}~\mathcal{P}_f^\alpha(\theta)\Bigr]|\Omega_\alpha\rangle\,,
\label{Bfree}
\end{equation}
with
\EQ
\mathcal{P}_f^\alpha(\theta)=\sum_{\beta\not=\alpha}\left[K_1(\theta)\,A^{\dagger}_{\alpha\beta}(-\theta)A^{\dagger}_{\beta\alpha}(\theta)+K_2(\theta)\sum_{\gamma\not=\alpha,\beta}A^{\dagger}_{\gamma\beta}(-\theta)A^{\dagger}_{\beta\alpha}(\theta)\right]\,,
\EN
\begin{align}
\label{K1}
&K_1(\theta)=(q-3)\frac{\sinh\bigl[\lambda(4i\pi/3-2\theta)\bigr]}{\sinh\bigl(2\lambda\theta\bigr)}\frac{\Gamma\bigl(-4\lambda/3+2\hat{\theta}+1\bigr)\Gamma\bigl(7\lambda/3-2\hat{\theta}\bigr)}{\Gamma\bigl(2\lambda/3-2\hat{\theta}+1\bigr)\Gamma\bigl(\lambda/3+2\hat{\theta}\bigr)}\,Q(\theta)\,,\\
\label{K2}
&K_2(\theta)=\frac{\sin\frac{2\pi\lambda}{3}}{\sin\frac{\pi\lambda}{3}}\frac{\sinh\bigl[\lambda(i\pi-2\theta)\bigr]}{\sinh\bigl(2\lambda\theta\bigr)}\frac{\sinh\bigl[\lambda(4i\pi/3-2\theta)\bigr]}{\sinh\bigl[\lambda(-2i\pi/3+2\theta)\bigr]}\frac{\Gamma\bigl(-4\lambda/3+2\hat{\theta}+1\bigr)\Gamma\bigl(7\lambda/3-2\hat{\theta}\bigr)}{\Gamma\bigl(2\lambda/3-2\hat{\theta}+1\bigr)\Gamma\bigl(\lambda/3+2\hat{\theta}\bigr)}\,Q(\theta)\,,\\
\label{Q_int}
&\hspace{1cm}Q(\theta)=\exp\Biggl[-\int_{0}^{\infty}\frac{dt}{t}\frac{\text{e}^{-2t}\sinh\bigl(\frac{5t}{6}-\frac{t}{2\lambda}\bigr)-\sinh\bigl(\frac{3t}{2}-\frac{t}{2\lambda}\bigr)}{2\cosh t\sinh\bigl(\frac{t}{2\lambda}\bigr)\cosh\bigl(\frac{t}{2}\bigr)}\sinh\Bigl(t-\frac{2\theta t}{i\pi}\Bigr)\Biggr],
\end{align}
where $\hat{\theta}\equiv\frac{\lambda\theta}{i\pi}$; the integral in (\ref{Q_int}) is again convergent for $1/2\leq\lambda<1$. In this case the residue at $\theta=0$ is non-zero and gives\footnote{There appears to be a typo in eq.~(48) of \cite{Chim}. In particular it does not reproduce $\tilde{g}_f^2\bigl(\frac{3}{4}\bigr)=1$ for the Ising model (q=2).}
\EQ
\tilde{g}_f^2=\frac{i}{2}\text{Res}_{\theta=0}K_1(\theta)=\frac{i}{2}\text{Res}_{\theta=0}K_2(\theta)=\frac{(3-q)}{4}\frac{\sin\frac{4\pi\lambda}{3}}{\lambda}\frac{\Gamma\bigl(1-\frac{4\lambda}{3}\bigr)\Gamma\bigl(\frac{7\lambda}{3}\bigr)}{\Gamma\bigl(1+\frac{2\lambda}{3}\bigr)\Gamma\bigl(\frac{\lambda}{3}\bigr)}Q(0)\,,
\EN
\EQ
\mathcal{S}_f^\alpha=1+\tilde{g}_f\sum_{\beta\not=\alpha}A^{\dagger}_{\beta\alpha}(0)\,.
\EN
We also quote the boundary cross-unitarity conditions
\begin{align}
\label{crossing_free}
&K_1(\theta)=\bigl[S_{\alpha\alpha}^{\beta\beta}(2\theta)+(q-2)S_{\alpha\alpha}^{\beta\gamma}(2\theta)\bigr]K_1(-\theta),\\
&K_2(\theta)=\bigl[S_{\alpha\gamma}^{\beta\beta}(2\theta)+(q-3)S_{\alpha\gamma}^{\beta\delta}(2\theta)\bigr]K_2(-\theta).
\end{align}

\section{Partition functions and final results}
In principle, the knowledge of the bulk and boundary amplitudes should allow the study of partition functions on the strip for any $R$ through the boundary version \cite{LMSS} of the thermodynamic Bethe ansatz (TBA) \cite{TBA}. In practice, however, the very non-trivial structure of Potts field theory seriously complicates the task\footnote{See \cite{DPT} for the state of the art of TBA in the Potts model.}. More pragmatically, here we plug the explicit expressions for $|B_\alpha\rangle$ and $|B_f\rangle$ into (\ref{Z_ud}) and exploit the fact that the states (\ref{states}) are eigenstates of the Hamiltonian $H$ with eigenvalues $m\sum_{i=1}^n\cosh\theta_i$, $m$ being the mass of the kinks. This leads to a large $R$ expansion for the partition functions for which we compute below the terms coming from one- and two-kink states.

\begin{figure}[t]
\begin{center}
\includegraphics[width=9cm]{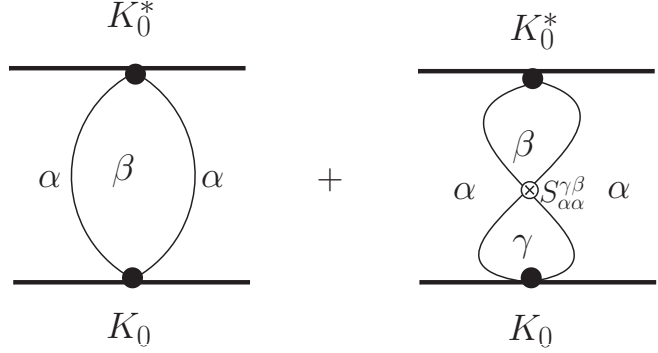}
\caption{Diagrammatic interpretation of the two contributions entering the matrix element (\ref{matrix_element}).}
\label{fig_expansion}
\end{center}
\end{figure}

Since we work in the kink basis, the partition functions we obtain in this way are those above $J_c$, that we denote $\tilde{Z}_{ud}^{lr}$, keeping the notation ${Z}_{ud}^{lr}$ for those below $J_c$. As an illustration, for $\tilde{Z}_{\alpha\alpha}^{ff}$ expansion of the boundary state leads to
\EQ
\tilde{Z}_{\alpha\alpha}^{ff}(R)=1+\frac{1}{4}\int_{-\infty}^{\infty}\frac{\text{d}\theta\text{d}\theta'}{(2\pi)^2}K_0^{*}(\theta')K_0(\theta)\text{e}^{-2mR\cosh\theta}\sum_{\beta,\gamma\not=\alpha}M_{\alpha\beta\gamma\alpha}(\theta,\theta')+O(\text{e}^{-4mR})\,,
\EN
\begin{align}
 M_{\alpha\beta\gamma\alpha}(\theta,\theta')&\equiv\langle\Omega_{\alpha}|A_{\alpha\beta}(\theta')A_{\beta\alpha}(-\theta')A^{\dagger}_{\alpha\gamma}(-\theta)A^{\dagger}_{\gamma\alpha}(\theta)|\Omega_{\alpha}\rangle\nonumber\\
&=(2\pi)^2\bigl[\delta(\theta'-\theta)\bigr]^2\delta_{\beta\gamma}+(2\pi)^2\bigl[\delta(\theta'+\theta)\bigr]^2 S_{\alpha\alpha}^{\gamma\beta}(2\theta')\,;\label{matrix_element}
\end{align}
the last equality follows from formal use of (\ref{FZ2kinks}) and can be associated to the diagrams shown in Fig.~\ref{fig_expansion}. If $L\to\infty$ denotes the horizontal size of the system, the squared delta functions in (\ref{matrix_element}) admit the usual regularization\footnote{It is important to stress that, as observed for other models in \cite{LMSS} (see also \cite{BPT05}), contributions to (\ref{Z_ud}) coming from states with more than two particles produce singularities whose regularization depends in general on the interaction. This is what makes difficult the determination of additional terms in the large $R$ expansion within the approach we are following.}
\EQ
\bigl[\delta(\theta'\pm\theta)\bigr]^2\rightarrow\delta(\theta'\pm\theta)\,\frac{mL}{2\pi}\,\cosh\theta\,;
\EN
free vertical boundary conditions have been imposed making no selection on the kink states which propagate between the horizontal boundaries.
Exploiting boundary cross-unitarity (\ref{crossing_fixed}) and real analiticity $K_0(-\theta)=K^{*}_0(\theta)$, $\theta\in\mathbb R$, we then obtain

\bea
\tilde{Z}_{\alpha\alpha}^{ff}(L,R)&=&1+(q-1)\,mL\,F_{\alpha\alpha}(R)+O(\text{e}^{-4mR})\,,\hspace{.5cm}mL\gg 1\,,\label{Z_aa}\\
F_{\alpha\alpha}(R)&=&\int_{0}^{\infty}\frac{\text{d}\theta}{2\pi}~\cosh\theta~|K_0(\theta)|^2~\text{e}^{-2mR\cosh\theta}\,.
\eea
Similarly one finds
\bea
\label{Zff}
\tilde{Z}_{ff}^{ff}(L,R) &=& q\left[1+(q-1)\,mL\,\Bigl(\tilde{g}_f^2~\text{e}^{-mR}+F_{ff}(R)\Bigr)\right]+O(\text{e}^{-3mR})\,,\hspace{.5cm}mL\gg 1\,,\\
F_{ff}(R) &=& \int_{0}^{\infty}\frac{\text{d}\theta}{2\pi}~\cosh\theta~\Bigl(|K_{1}(\theta)|^2+(q-2)|K_2(\theta)|^2\Bigr)~\text{e}^{-2mR\cosh\theta}\,,
\label{F_ff}\\ \nonumber \\
\label{Zaf}
\tilde{Z}_{\alpha f}^{ff}(L,R) &=& 1+(q-1)\,mL\,F_{\alpha f}(R)+O(\text{e}^{-4mR})\,,\hspace{.5cm}mL\gg 1\,,\\
F_{\alpha f}(R) &=& \int_{0}^{\infty}\frac{\text{d}\theta}{2\pi}~\cosh\theta\,\mbox{Re}\bigl[K_0^{*}(\theta)K_{1}(\theta)\bigr]~\text{e}^{-2mR\cosh\theta}\,,
\label{F_alphaf}
\eea
and $\tilde{Z}_{f\alpha}^{ff}=\tilde{Z}_{\alpha f}^{ff}$. The partition functions $\tilde{Z}_{ud}^{\beta\alpha}$ with fixed vertical boundary conditions are obtained taking off from $\tilde{Z}_{ud}^{ff}$ the contribution of the states which are not of the form (\ref{states}) with $\alpha_1=\alpha$ and $\alpha_{n+1}=\beta$.

From these results we obtain for the mean number of crossing clusters in the scaling limit above $p_c$ the universal result
\EQ
\tilde{\bar{N}}_v(L,R)=\lim_{q\rightarrow 1}\partial_q\log\frac{\tilde{Z}_{\alpha\alpha}^{ff}\,\tilde{Z}_{ff}^{ff}}{\tilde{Z}_{f\alpha}^{ff}\,\tilde{Z}_{\alpha f}^{ff}}\sim 1+mL\left[\Phi(R)+O(\text{e}^{-3mR})\right]\,,\hspace{.6cm}mL\gg 1\,,
\label{Nv+}
\EN
\EQ
\Phi(R)=A\,\text{e}^{-mR}+[F_{ff}(R)+F_{\alpha\alpha}(R)-2F_{\alpha f}(R)]_{q=1}\,,
\label{phi}
\EN
where $A=\tilde{g}^2_f|_{q=1}$ reduces to (\ref{A}). The additive term 1 in (\ref{Nv+}) is produced by the overall factor $q$ in (\ref{Zff}) and accounts for the contribution of the infinite cluster in the limit $R\to\infty$. Notice that any normalization of the boundary states other than the one we used would anyway cancel in the combination of partition functions in  (\ref{Nv+}).

In order to determine the mean number of crossing clusters below $p_c$ we have to use the duality \cite{Potts} of the Potts model to connect the partition functions (\ref{Zaa}-\ref{Zfa}) below $J_c$ to the partition functions $\tilde{Z}_{ud}^{lr}$ above $J_c$ we have computed. Duality maps free boundary conditions into fixed boundary conditions and vice versa (see e.g. \cite{Wu_duality}). For our present purpose of counting the vertically crossing clusters, it is useful to observe that fixing the spins to the color $\alpha$ on both vertical sides, rather than leaving them free, has the only effect that the clusters touching at least one vertical side are not counted. Below $p_c$, where all clusters are finite with a mean linear extension of order $\xi$, such a boundary term does not affect $\bar{N}_v$, which is extensive in $L$ in the limit $L/\xi\to\infty$ we are considering. So we can use (\ref{Nc}) with the replacement $ff\to\alpha\alpha$ in the vertical boundary conditions, and use duality\footnote{In principle $Z_{ff}^{\alpha\alpha}$ could be mapped into a linear combination of $\tilde{Z}_{\alpha\alpha}^{ff}$ and $\tilde{Z}_{\alpha\beta}^{ff}$, $\alpha\neq\beta$. However, it follows from (\ref{Pa}) that the latter partition function vanishes identically on the infinitely long strip; at large $L$ it is suppressed as $e^{-mL}$.} to obtain in the scaling limit below $p_c$
\EQ
\bar{N}_v(L,R)\sim\left[\lim_{q\rightarrow 1}\partial_q\log\frac{\tilde{Z}_{\alpha\alpha}^{ff}\,\tilde{Z}_{ff}^{ff}}{\tilde{Z}_{f\alpha}^{ff}\,\tilde{Z}_{\alpha f}^{ff}}\right]_{\text{extensive part}}=mL\left[\Phi(R)+O(\text{e}^{-3mR})\right]\,,\hspace{.6cm}mL\gg 1\,.
\label{spanning_clusters_result}
\EN
A different derivation of this result is given in the Appendix.

The functions $F_{ff}(R)|_{q=1}$ and $F_{\alpha f}(R)|_{q=1}$ in (\ref{phi}) are well defined in spite of the poles at $\theta=0$ in the amplitudes $K_1(\theta)$ and $K_2(\theta)$ contained in the integrands in (\ref{F_ff}) and (\ref{F_alphaf}). If the convergence of (\ref{F_alphaf}) simply follows from $K_0(0)=0$, the case of $F_{ff}$ is more subtle. Consider indeed the Laurent expansions $K_i(\theta)={a_{-1}}/{\theta}+a_0^{(i)}+a_1^{(i)}\,\theta+\dots$, $i=1,2$.
If $\theta\in\mathbb R$, due to the relation $K_i^{*}(\theta)=K_i(-\theta)$ the coefficients $a_{2k-1}^{(i)}$ are purely imaginary and the coefficients $a_{2k}^{(i)}$ are instead real for all non-negative integers $k$; this in turn implies that the combination $|K_1(\theta)|^2 -|K_2(\theta)|^2$ entering $F_{ff}(R)|_{q=1}$ does not contain any double or single pole at $\theta=0$.

\begin{figure}[t]
\hspace{1.2cm}
\begin{minipage}[hb]{.4\textwidth}
\begin{center}
\includegraphics[width=7.5cm]{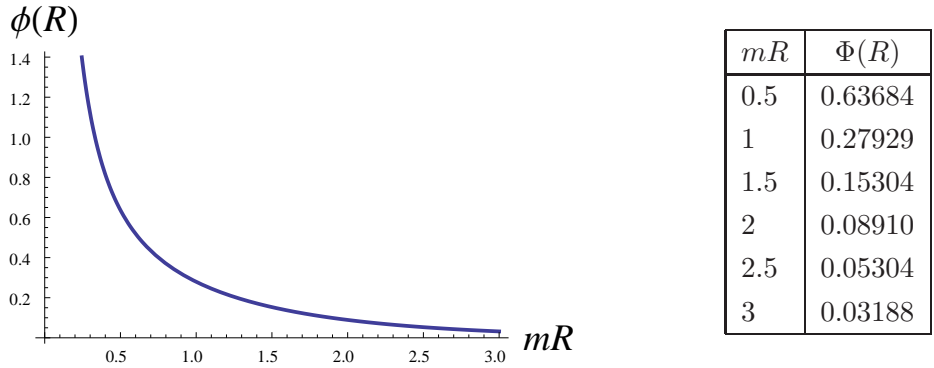}
\end{center}
\end{minipage}
\qquad\quad
\begin{minipage}[hb]{.4\textwidth}
\begin{center}
\begin{tabular}{|l|c|}
\hline$mR$ & $\Phi(R)$ \\
\hline
0.5 & 0.63684 \\
1 & 0.27929 \\
1.5 & 0.15304  \\
2 & 0.08910\\
2.5 & 0.05304 \\
3  & 0.03188\\
\hline
\end{tabular}
\end{center}
\end{minipage}
\caption{Plot of the function (\ref{phi}); few values are given in the table.}
\label{fig_plot}
\end{figure}

The function (\ref{phi}) is plotted in Fig. \ref{fig_plot}, where some numerical values are also listed. Since (\ref{spanning_clusters_result}) is extensive in $L$ for any $R$, it is tempting to check what our large $R$ result gives in the conformal limit $mR\rightarrow 0$, for which the result $\bar{N}_v\sim(\sqrt{3}/4)L/R=(0.433..)L/R$, $L\gg R$,
is known from \cite{Cardy00}. Using the large $\theta$ limits $K_0\to\text{e}^{i\pi/3}$, $K_1\to -2\text{e}^{i\pi/3}$ and $K_2\to -\sqrt{3}\text{e}^{i\pi/6}$ at $q=1$, (\ref{spanning_clusters_result}) gives\footnote{Given $F(y)=\int_{0}^{\infty}\text{d}x~\cosh x~\text{e}^{-y\cosh x}f(x)$, with $\lim_{x\to\infty}f(x)=\alpha$, we have $F(y)\to\alpha/y$ for $y\to 0$.} $3/(2\pi)(L/R)=(0.477..)L/R$, with a $10\%$ deviation from the exact result suggesting that (\ref{phi}) may still provide a good approximation for $mR$ of order 1.

We already explained how (\ref{P}) follows from (\ref{spanning_clusters_result}). The same result also follows from the observation that a lattice configuration with a vertical crossing is mapped onto a dual lattice configuration without horizontal crossings, and vice versa \cite{Wu}, so that\footnote{In the continuum limit at $p_c$ (\ref{hv}) reproduces the known relation $P_v+P_h=1$ \cite{LPPS-A}.}
\EQ
P_v=1-\tilde{P}_h=\lim_{q\to 1}\tilde{Z}_{ff}^{\alpha\beta}\,.
\label{hv}
\EN
The partition function $\tilde{Z}_{ff}^{\alpha\beta}$ is obtained picking up in (\ref{Zff}) only the contributions of the states compatible with the vertical boundary conditions $\alpha\beta$. In particular, since we are no longer summing over $\alpha$ and $\beta$, the one-kink contribution in $e^{-mR}$ now appears with multiplicity one rather than $q(q-1)$, and this gives (\ref{P}) back\footnote{Notice that our normalization of the boundary states ensures the conditions
$\lim_{q\to 1}\tilde{Z}_{ff}^{ff}=\lim_{q\to 1}\tilde{Z}_{\alpha\alpha}^{\alpha\alpha}=1$ required for percolation.}. Concerning the two-kink contribution, only the term containing $|K_2|^2$ survives now, again without the prefactor $q(q-1)$; $|K_2|^2$ contains a singularity of the form $4\tilde{g}_f^4/\theta^2$ at $\theta=0$ which has to be subtracted\footnote{The result obtained in \cite{BPT05} for simpler models (with purely transmissive scattering) by a TBA analysis amounts to such a subtraction.} because produced by the propagation of states created by $A^\dagger_{\alpha\gamma}(0)A^\dagger_{\gamma\beta}(0)$ which, as already observed, are not compatible with (\ref{FZ1kinks}) and (\ref{fermionic_kink}). Hence, the contribution of order $e^{-2R/\xi}$ to be added to (\ref{P}) is
\EQ
U(L,R)=-mL \int_{0}^{\infty}\frac{\text{d}\theta}{2\pi}~\cosh\theta~\Bigl(|K_2(\theta)|_{q=1}^2\,\text{e}^{-2mR\cosh\theta}-\frac{4A^2}{\sinh^2\theta}\,\text{e}^{-2mR}\Bigr)\,.
\label{U}
\EN
The replacement $1/\theta^2\to1/\sinh^2\theta$, relevant for the convergence of the integral, comes from the fact that $m\int d\theta\cosh\theta=\int dp$, and in the momentum variable $p$ the function $|K_2|^2$ diverges as $1/p^2$.

The vertical crossing probability above $p_c$ is $\tilde{P}_v=1-\lim_{q\to 1}\tilde{Z}_{\alpha\beta}^{ff}$. We already observed that $\tilde{Z}_{\alpha\beta}^{ff}$ vanishes exponentially at large $L$, in agreement with the expectation that, due to the presence of an infinite cluster, above $p_c$ the crossing probability tends to 1 as we enlarge the window. More precisely, duality gives
\bea
\tilde{P}_v(L,R)&=& 1-P_h(L,R)\nonumber\\
&\sim & 1-A~\frac{R}{2\xi}\,\text{e}^{-L/2\xi}-U(R,L)\,,\hspace{1cm}R\gg\xi\,,\hspace{.3cm}L\gtrsim R\,,
\label{Ptilde}
\eea
where the last line takes (\ref{xi}) into account and holds at order $e^{-L/\xi}$.

\vspace{1cm} 
\noindent
\textbf{Acknowledgments.} We thank J. Cardy for a discussion about eq.~(\ref{hv}).

\begin{figure}[t]
\begin{center}
\includegraphics[height=6cm]{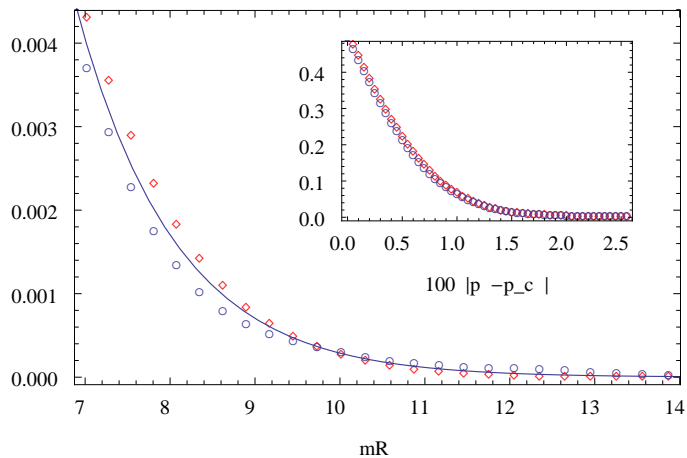}
\caption{The inset shows Monte Carlo data from \cite{WYNH} for the crossing probability below $p_c$ (diamonds) and for the complement to 1 of the crossing probability above $p_c$ (circles); the data refer to bond percolation on the square lattice with $L=R=256$. The tails are plotted against $mR$ using $m=m_0|p-p_c|^{4/3}$, with $m_0=5.8$. The continuous curve is the result (\ref{P}), (\ref{A}), (\ref{Ptilde}), i.e. $A\,mR\,e^{-mR}$.}
\label{cross_tails}
\end{center}
\end{figure}

\vspace{1cm}
\noindent
\textbf{Note added.} We learned from a referee that a scaling analysis of Monte Carlo data for $P_v(L,R)$ in terms of a single scaling variable was performed in \cite{WYNH} and further discussed in \cite{PM,WH}. It is relevant for the present paper that the data of \cite{WYNH} allow a comparison with our results. The inset of Fig.~\ref{cross_tails} shows the data of \cite{WYNH} for the crossing probability in bond percolation on the square lattice of size $L=R=256$ lattice units; they satisfy the duality relation (\ref{hv}) for the crossing probability above and below $p_c$ (which for $R=L$ specializes to $P_v=1-\tilde{P}_v$) up to discrepancies to be ascribed to a mixing of finite size effects, corrections to scaling and statistical errors. In principle comparison of the data to (\ref{P}), (\ref{A}) and (\ref{Ptilde}) allows to fit the value of the only unknown parameter, i.e. the non-universal amplitude $m_0$ entering the relation $m=m_0|p-p_c|^\nu$, $\nu=4/3$. We fit $m_0\approx 5.7$ from the tail of the crossing probability below $p_c$, and $m_0\approx 5.9$ from the tail above $p_c$; consider that for $R=256$ and $mR$ around 10, $\xi$ is around 25 below $p_c$ and around 12 above, so that the analysis is almost certainly affected by non-negligible corrections to scaling. On the other hand, the value $2\xi_{2nd}^0\approx 0.37$ was measured in \cite{DAS} for the amplitude of the second moment correlation length in bond percolation on the square lattice below $p_c$, and it is known from \cite{DVC} that $m_0=a/\xi_{2nd}^0\approx 5.4\,a$, with $a$ equal 1 up to corrections that are not expected to exceed few percents. A comparison between the data for the tails and (\ref{P}), (\ref{A}) is given in Fig.~\ref{cross_tails}; the subleading term (\ref{U}) is always very small and totally negligible in the range of $mR$ shown in the figure. Putting all together, our conclusion is that the data\footnote{Aspect ratios $L/R>1$ were also analyzed in \cite{WYNH}, but for these cases the range of $mL$ covered by the data is not large enough to allow comparison with (\ref{Ptilde}).} of \cite{WYNH} are consistent with the results of this paper within the numerical uncertainties; an unambiguous verification will require simulations expressly targeting the tails on larger lattices.
 
We are very grateful to H. Watanabe and C.-K. Hu for providing us with the data of \cite{WYNH}, and to the referee for bringing references \cite{WYNH,PM,WH} to our attention and for noticing that the constant $A$, that we originally quoted in the form $\left(\frac{3}{2}(2-\sqrt{3})\right)^{1/2}$, can equivalently be written as in  (\ref{A}).

\vspace{.3cm}
\section*{Appendix}
Let us denote by $\phi_{ab}(x)$ the field \cite{Cardy89} whose insertion at point $x$ on the boundary changes the boundary condition from $a$ to $b$; of course $\phi_{aa}$ coincides with the identity $I$. If $x_1,\ldots,x_4$ are the coordinates of the corners of the rectangle starting from the left upper corner and moving clockwise, we have\footnote{At $J_c$ analogous relations were used in \cite{Cardy92,Cardy00}.}
\bea
& G_{lurd}\equiv\langle\phi_{lu}(x_1)\phi_{ur}(x_2)\phi_{rd}(x_3)\phi_{dl}(x_4)\rangle_{J\leq J_c}=Z_{ud}^{lr}/Z_{ll}^{ll}\,,\\
& \tilde{G}_{lurd}\equiv\langle\phi_{lu}(x_1)\phi_{ur}(x_2)\phi_{rd}(x_3)\phi_{dl}(x_4)\rangle_{J^*\geq J_c}=\tilde{Z}_{ud}^{lr}/\tilde{Z}_{ll}^{ll}\,.
\label{corr}
\eea
The fields $\phi_{ab}$ obey the natural operator product expansion \cite{Cardy92}
\EQ
\phi_{\alpha f}\cdot\phi_{f\beta}=\delta_{\alpha\beta}\,I+c\,(1-\delta_{\alpha\beta})\mu_{\alpha\beta}+\cdots\,,
\label{bc1}
\EN
that we write symbolically omitting the coordinate dependence, and using the notation $\mu_{\alpha\beta}=\phi_{\alpha\beta}$ for the kink field which switches between fixed boundary conditions with different colors. The field $\mu_{\alpha\beta}(x)$ is dual to the Potts spin field $\sigma_\alpha(x)=q\delta_{s(x),\alpha}-1$, and the relation
\EQ
\langle\sigma_\alpha(x)\sigma_\beta(y)\rangle_{J\leq J_c}=(q\delta_{\alpha\beta}-1)\,\langle\mu_{\gamma\delta}(x)\mu_{\delta\gamma}(y)\rangle_{J^*\geq J_c}
\label{sigmamu}
\EN
holds (see e.g. \cite{duality}). Since duality exchanges fixed and free boundary conditions, we then write the dual of (\ref{bc1}) as
\EQ
\phi_{f\alpha}\cdot\phi_{\alpha f}=I+c'\,\sigma_\alpha+\cdots\,,
\label{bc2}
\EN
with $c'$ a new structure constant. For the boundary correlators $G_{lurd}$, we have simple duality relations like $G_{fff\alpha}=\tilde{G}_{\alpha\alpha\alpha f}$, but also non-trivial ones like 
\bea
& G_{f\alpha f\alpha}=a_1\,\tilde{G}_{\alpha f\alpha f}+a_2\,\tilde{G}_{\alpha f\beta f}\,,
\label{d1}\\
& G_{f\alpha f\beta}=a_3\,\tilde{G}_{\alpha f\alpha f}+a_4\,\tilde{G}_{\alpha f\beta f}\,,
\label{d2}
\eea
where $\alpha\neq\beta$. In order to determine the coefficients $a_1,\ldots,a_4$  we use (\ref{bc1}) and (\ref{bc2}) to take the limits $x_1\to x_2$ and $x_3\to x_4$ on both sides,  take (\ref{sigmamu}) into account to equate the coefficients of the two-point functions we are left with, and obtain
\EQ
a_1=a_3=1\,,\hspace{1cm}a_2=(q-1)(c'/c)^2\,,\hspace{1cm}a_4=-(c'/c)^2\,.
\label{ai}
\EN
Since $P_v=1-\lim_{q\to 1}Z_{\alpha\beta}^{ff}=1-\lim_{q\to 1}\left[a_3+a_4\tilde{Z}_{ff}^{\alpha\beta}\right]$, comparison with (\ref{hv}) and (\ref{ai}) gives $(c'/c)^2=1$ at $q=1$.
Putting all together, the combination of partition functions in (\ref{Nc}) can be written as
\EQ
R=\frac{Z_{ff}^{ff}Z_{\alpha\alpha}^{ff}}{Z_{\alpha f}^{ff}Z_{f\alpha}^{ff}}=\frac{G_{f\alpha f\alpha}}{G_{f\alpha f f}\,G_{fff\alpha}}=\frac{\tilde{G}_{\alpha f\alpha f}+a_2\,\tilde{G}_{\alpha f\beta f}}{\tilde{G}_{\alpha f\alpha\alpha}\,\tilde{G}_{\alpha\alpha\alpha f}}=\frac{\tilde{Z}_{\alpha \alpha}^{\alpha\alpha}\left[\tilde{Z}_{ff}^{\alpha\alpha}+a_2\,\tilde{Z}_{ff}^{\alpha\beta}\right]}{\tilde{Z}_{f\alpha}^{\alpha\alpha}\,\tilde{Z}_{\alpha f}^{\alpha\alpha}}\,,
\label{R}
\EN
with $a_2=q-1+O((q-1)^2)$. Now it is not difficult to use our expressions for the partition functions $\tilde{Z}_{ud}^{lr}$ to check that $\lim_{q\to 1}\partial_q\log R$ gives the r.h.s. of (\ref{spanning_clusters_result}).



\begin{thebibliography}{99}
\bibitem{LPPS-A} R. Langlands, C. Pichet, P. Pouliot and Y. Saint-Aubin, J. Stat. Phys. 67 (1992) 553.
\bibitem{Cardy92} J. Cardy, J. Phys. A 25 (1992) L201.
\bibitem{Watts} G. Watts, J. Phys. A 29 (1996) L363.
\bibitem{LPS-A} R. Langlands, P. Pouliot and Y. Saint-Aubin, Bull. Amer. Math. Soc. (N.S.) 30 (1994) 1.
\bibitem{S.Smirnov} S. Smirnov, C.R. Acad. Sci. Paris, t. 333, S\'erie I (2001) 339.
\bibitem{Cardy00} J. Cardy, Phys. Rev. Lett. 84 (2000) 3507.
\bibitem{CardyLP} J. Cardy, Lectures on Conformal Invariance and Percolation, arXiv:math-ph/0103018.
\bibitem{DVC} G. Delfino and J. Cardy, Nucl. Phys. B 519 (1998) 551.

G. Delfino, J. Viti and J. Cardy, J. Phys. A 43 (2010) 152001.
\bibitem{BW} L. Berlyand and J. Wehr, J. Phys. A 28 (1995) 7127.
\bibitem{HA} J.-P. Hovi and A. Aharony, Phys. Rev. E 53 (1996) 235.
\bibitem{NZ} M. Newman and R. Ziff, Phys. Rev. Lett. 85 (2000) 4104; Phys. Rev. E 64 (2001) 16706. 
\bibitem{ONS} P. de Oliveira, R. N\'obrega and D. Stauffer, J. Phys. A 37 (2004) 3743.
\bibitem{Vasilyev} O.A. Vasilyev, Phys. Rev. E 72 (2005) 036115.
\bibitem{Potts} R.B. Potts, Proc. Cambr. Phil. Soc. 48 (1952) 106.
\bibitem{Wu} F.Y. Wu, Rev. Mod. Phys. 54 (1982) 235.
\bibitem{FK}  C.M. Fortuin and P.W. Kasteleyn, J. Phys. Soc. Jpn. Suppl. 26 (1969) 11; Physica 57 (1972) 536.
\bibitem{CZ} L. Chim and A.B. Zamolodchikov, Int. J. Mod. Phys. A 7 (1992) 5317.
\bibitem{Smirnov}  F. Smirnov, Form Factors in Completely Integrable Quantum Field Theory, World Scientific (1992).
\bibitem{GZ94} S. Ghoshal and A.B. Zamolodchikov, Int. J. Mod. Phys. A 9 (1994) 3841.
\bibitem{BPT07} Z. Bajnok, L. Palla and G. Takacs, Nucl. Phys. B 772 (2007) 290.
\bibitem{DPTW00} P. Dorey, M. Pillin, R. Tateo and G. Watts, Nucl. Phys. B 594 (2000) 625.
\bibitem{Chim} L. Chim, J. Phys. A: Math Gen. 28 (1995) 7039.
\bibitem{LMSS} A. Le Clair, G. Mussardo, H. Saleur and S. Skorik, Nucl. Phys. B 453 (1995) 581.
\bibitem{TBA} Al. Zamolodchikov, Nucl. Phys. B 342 (1990) 695.
\bibitem{DPT} P. Dorey, A. Pocklington and R. Tateo, Nucl. Phys. B 661 (2003) 464.
\bibitem{BPT05} Z. Bajnok, L. Palla and G. Takacs, Nucl. Phys. B 716 (2005) 519.
\bibitem{Wu_duality} F.Y. Wu, Phys. Lett. A 228 (1997) 43.

\bibitem{Cardy89} J. Cardy, Nucl. Phys. B 324 (1989) 581.
\bibitem{duality} G. Delfino and J. Viti, Nucl. Phys. B 852 (2011) 149.
\bibitem{WYNH} H. Watanabe, S. Yukawa, N. Ito and C.-K. Hu, Phys. Rev. Lett. 93 (2004) 190601.
\bibitem{PM} G. Pruessner and N.R. Moloney, Phys. Rev. Lett. 95 (2005) 258901.
\bibitem{WH} H. Watanabe and C.-K. Hu, Phys. Rev. Lett. 95 (2005) 258902; 
Phys. Rev. E 78 (2008) 041131.
\bibitem{DAS} D. Daboul, A. Aharony and D. Stauffer, J. Phys. A 33 (2000) 1113.


\end{thebibliography}
\end{document}